\documentstyle[epsfig,aps]{revtex}

\begin{document}

\title{Nonlinear analysis of the shearing instability in granular gases}
\author{R. Soto, M. Mareschal\\
{\em \small CECAM, ENS-Lyon, 46 All\'ee d'Italie, 69007 Lyon, France}\\
M. Malek Mansour\\
{\em \small Universit\'e Libre de Bruxelles, Campus Plaine, Bvd. du Triomphe,
CP 231,
1050  Bruxelles,
Belgium}}
\maketitle

\begin{abstract}

It is known that a finite-size homogeneous granular fluid develops an
hydrodynamic-like instability when dissipation crosses a threshold
value. This instability is  analyzed in terms of modified hydrodynamic
equations: first,  a source term is added to the energy equation
which accounts for the energy dissipation at collisions and the
phenomenological Fourier law is generalized according to previous
results. Second, a rescaled  time formalism is introduced that maps
the homogeneous cooling state into a nonequilibrium steady state. A
nonlinear stability analysis of the resulting  equations is done which
predicts the appearance of flow patterns.  A stable
modulation of density and temperature is produced that does not lead to
clustering. Also a global decrease of the temperature is obtained,
giving rise to a decrease of the collision frequency and dissipation
rate. Good agreement with molecular dynamics simulations of inelastic
hard disks is found for low dissipation.

\end{abstract}

\pacs{
45.70.-n, 
45.70.Mg, 
47.20.-k, 
47.20.Ky  
}

\section{Introduction}

The understanding of the dynamics of granular fluids is crucial
for various industrial processes. This has led to many investigations
where theory, experiments and simulations are used in order to construct a
predictive theory
\cite{campbell90,jaeger92,melo94,umbanhowar96,jaeger96,olafsen98a,falcon99,wildman99}. 
There is hope that for moderate
densities and slightly dissipative grains, grain dynamics may be
described on large scales by using fluid hydrodynamics with slight
modifications. 

 One way to guess which changes to make in standard hydrodynamics is to
use the tools developed  by statistical mechanics in order to derive
hydrodynamic equations. This is justified by the fact that a simple
grain model, the inelastic hard sphere model, has been shown to
reproduce most of the phenomena occuring in granular systems: in some
sense, it has proven to contain the essential ingredients necessary to
predict the peculiar physics observed
\cite{goldhirsch93,mcnamara96,muller97,zamankhan97b,grossman97,rapaport98}.

 The scheme used in the study of nonequilibrium fluids can then be
extended to granular fluids: ``microscopic'' simulations permit to compute
the equation of state and values for transport coefficients which are
then fed into the guessed macroscopic equations. Comparison between
direct nonequilibrium simulations of microscopic and macroscopic models
allows then to test the validity of the proposed macroscopic
equations. This approach was used recently by two of us to investigate
 heat transport (heat being identified with the kinetic energy
associated with the grains' motion) and it has been shown that Fourier's
law has to be generalized with a density gradient term appearing in the
expression for the heat flux \cite{sotomar99}.

In the Inelastic Hard Sphere model (IHS), grains are spherical hard
particles with only translational degrees of freedom. The energy
dissipation is included through a restitution coefficient $r$ lower than
one.  As for hard spheres, the collision is an instantaneous event and
the grain velocities after a
 collision are given by
\begin{eqnarray} {\bf v}_1' &=& {\bf v}_1 + \frac{1}{2} (1+r)
\left[\widehat{\bf
n} \cdot ({\bf v}_2 -{\bf v}_1)\right] \widehat{\bf n}\\ {\bf v}_2' &=& {\bf
v}_2 - \frac{1}{2} (1+r) \left[\widehat{\bf n} \cdot ({\bf v}_2 -{\bf v}_1)
\right] \widehat{\bf n}
\end{eqnarray}
where $\widehat{\bf n}$ is the unit vector pointing from the center of particle
1 towards the center of particle 2. It is convenient to define the
dissipation coefficient $q=(1-r)/2$ which vanishes when collisions are
elastic. In what follows units are chosen such that the disk diameter
$\sigma$ and the particles masses are set to one.

The IHS model, like the elastic hard sphere model, does not have an
intrinsic energy scale.
This means that two systems with the same initial configuration and with
the particle velocities of one system being equal to those of the other
multiplied by one scaling factor will follow the same trajectory but at
different speeds.
 This lack of an intrinsic energy scale also implies
a simple temperature dependence for all hydrodynamic quantities which may
be found by simple dimensional analysis.

When a fluidized granular medium is let to evolve freely in a box with
periodic boundary conditions, the energy decreases continuously in time and
the system remains homogeneous. This non-steady state  is called the
Homogeneous Cooling State (HCS) and it is the reference state from which
perturbations are studied:  it is analogous to the thermodynamic
equilibrium state for elastic systems \cite{mcnamara96,brey96a,noije98c}.
In the IHS model this state is particularly simple and the energy decreases
obeying the Haff's law \cite{haff83}
\begin{equation}
E(t) = \frac{E(0)}{\left(1+t/t_0\right)^2}
\end{equation}
 In order to avoid the cooling of the system, the simulations are
made at constant energy: at every collision between the grains, the
dissipated energy is redistributed by scaling all the velocities. This is
equivalent to a time rescaling and can be  treated as such in the  equations
for the continuous model.  Using appropriate hydrodynamic equations, this
homogeneous state is predicted to be unstable under certain conditions of
density, system size and dissipativity
\cite{goldhirsch93,mcnamara96,brey96,brey96a,noije99,deltour97}.
Considering the dissipativity coefficient q as a bifurcation parameter,
while increasing q at constant density and number of grains, the system
first develops an instability characterized by two counterflows, the shearing
instability, and then either a clustering regime in which the density becomes
inhomogeneous or a vortex state where many small vortices develop
throughout the system.

For given values of total number of grains $N$ and number density $\rho=N/V$
the shearing instability appears when the dissipation coefficient is larger
than a critical value. In
the low density and large system limit,the critical value is given (in two
dimensions) by
\cite{jenkins85c}
\begin{equation}
\widehat{q} = \frac{\pi}{N \rho}
\end{equation}
Note that in the thermodynamic limit the system is always
unstable for any finite dissipation.

In this article we will study the shearing instability using a nonlinear
hydrodynamic approach. In Sec. \ref{hidros} we will develop a formalism that
allows us to treat the HCS as a non equilibrium steady state, and we will
present a stability analysis around the steady state. Next, in Sec.
\ref{nonlinear} we will present the nonlinear analysis of the instability,
obtaining expressions for the hydrodynamic fields beyond the instability.
It will also be shown that the presence of the instability modifies the
collision rate and energy dissipation rate. Finally, in Sec.
\ref{mdsim} we compare the predictions of the continuous model with
molecular dynamics simulations of inelastic hard disks. Conclusions are
presented in Sec.
\ref{conclusion}.

In what follows we will treat the two dimensional case, but the extension to
three dimensions is direct.


\section{Rescaled time formalism} \label{hidros}

Consider a two dimensional system composed of $N$ grains interacting with
the IHS model in a square box of size $L$ that has periodic boundary
conditions in both directions. Units are chosen such that Boltmann's
constant and the particle masses are set to one. Granular temperature
is defined 
analogously to the kinetic definition for classical fluids
\begin{equation}
T = \frac{1}{N}\sum_i \frac{1}{2} ({\bf v}_i-{\bf v})^2
\end{equation}
where ${\bf v}$ is the hydrodynamic velocity.

The shearing instability has been predicted by linear analysis of the
hydrodynamic equations for the IHS. Also an analysis of the first stages of
the nonlinear regime has been done in Ref. \cite{BreyNolinear}. As we shall
see, the relative simplicity of the model allows for a complete nonlinear
analysis as well.

The hydrodynamic equations for the low-density IHS are similar to the usual
hydrodynamic equations for fluids except that there is an energy sink term
and the heat flux has contributions from the density gradient
\cite{jenkins85c,jenkins83,brey96a,sotomar99}. The
equations read

\begin{eqnarray}
&&\frac{\partial\rho}{\partial t} + \nabla \cdot (\rho {\bf v}) = 0\\
&&\rho \left( \frac{\partial {\bf v}}{\partial t} + ({\bf v}\cdot\nabla){\bf v}\right) =
   - \nabla \cdot  I\!\! P\\
&&\rho \left( \frac{\partial T}{\partial t} + ({\bf v}\cdot\nabla)T\right) =
   - \nabla\cdot{\bf J} -  I\!\! P:\nabla{\bf v} -\omega
\label{eqhidro}
\end{eqnarray}
with the following constitutive equations
\begin{eqnarray}
 I\!\! P_{ij} &=& \rho T \delta_{ij} -\eta_0 \sqrt{T}
\left(\frac{\partial v_i}{\partial x_j} + \frac{\partial v_j}{\partial x_i} -
(\nabla\cdot{\bf v})\delta_{ij}\right)
\label{constrels}\\
{\bf J} &=& -k_0 \sqrt{T} \nabla T - \mu_0 \frac{T^{3/2}}{\rho} \nabla
\rho \nonumber\\
\omega &=& \omega_0 \rho^2 T^{3/2} \nonumber
\end{eqnarray}
where $\eta_0$, $k_0$, $\mu_0$, and $\omega_0$ do not depend on density or
temperature but on the dissipation coefficient $q$. In particular,
$\omega_0$ and $\mu_0$ vanish with $q$.

The total energy dissipation rate can be computed from the hydrodynamic
equations, obtaining
\begin{eqnarray}
\frac{dE}{dt} &=& \frac{d}{dt}\int (\rho T + \rho v^2/2)\, dV\\
              &=& -\int \omega \, dV
\label{energyrate}
\end{eqnarray}

In the HCS, where density and temperature are homogeneous and there is no
velocity field, the energy dissipation rate is
\begin{equation}
\frac{dE}{dt} = -V \omega_0 \rho_0^2 T^{3/2}\\
\end{equation}

This continuous cooling down has the disadvantage that any perturbation
analysis must
be done with respect to this non-steady state. To overcome this difficulty we
propose a differential time rescaling $ds = \gamma\, dt$ such that in this
rescaled time, energy is conserved.
The transformation corresponds to a continuous rescaling of all the
particle velocities such that the kinetic energy remains constant. This
rescaling does not, however, introduce new phenomena since as we have mentioned
the IHS does not have an intrinsic time scale, and thus a time rescaling
gives rise to the same phenomena viewed at different speed.

In order to keep the rescaled energy constant the
appropriate value of $\gamma$ is given by
\begin{equation}
\gamma = \sqrt{\frac{E(t)}{E(0)}}
\label{gamma}
\end{equation}

The rescaled time hydrodynamic fields transform as
\begin{eqnarray}
{\bf v}       &=& \gamma \widetilde{{\bf v}}
\label{tranform_s}\\
T             &=& \gamma^2 \widetilde{T}\\
 I\!\! P_{ij} &=& \gamma^2 \widetilde{ I\!\! P_{ij}}\\
{\bf J}       &=& \gamma^3 \widetilde{{\bf J}}\\
\omega        &=& \gamma^3 \widetilde{\omega}
\end{eqnarray}
where the tilde denotes the rescaled variable.

Due to the non-constant character of the time rescaling, extra source terms
appear in the hydrodynamic equations, which can be simplified using the
relation
\begin{equation}
\frac{1}{\gamma}\frac{d\gamma}{ds} = -\frac{1}{2 E(0)} \int
\widetilde{\omega}\, dV
\end{equation}

Suppressing the tildes everywhere, the equations now read:

\begin{eqnarray}
&&\frac{\partial \rho}{\partial s} + \nabla \cdot (\rho {\bf v}) = 0
\label{eqhidro_s}\\
&&\rho \left( \frac{\partial{\bf v}}{\partial s} + ({\bf v}\cdot\nabla){\bf v}\right) =
   - \nabla \cdot  I\!\! P + \frac{\rho {\bf v}}{2 E(0)}\int \omega \, dV
\nonumber\\
&&\rho \left( \frac{\partial T}{\partial s} + ({\bf v}\cdot\nabla)T\right) =
   - \nabla\cdot{\bf J} -  I\!\! P:\nabla{\bf v} -\omega
+ \frac{\rho T}{E(0)}\int \omega \, dV \nonumber
\end{eqnarray}

Note that the constitutive relations
(\ref{constrels}) remain unchanged under  time rescaling.

In the rescaled time, the HCS reduces to a non-equilibrium steady state with a
continuous energy supply that compensates the energy dissipation.  As there
is no energy scale, we fix the
reference temperature to be one, so that $E(0)=N$. The HCS is then
characterized by $\{\rho = \rho_H, \, T = 1, \, {\bf v} = 0 \}$.  To study
the stability of this state,
we first introduced the change of variables:
\begin{eqnarray}
\rho &=& \rho_H + \delta \rho\\
{\bf v}&=& \delta {\bf v}\\
T &=& 1 + \delta T
\end{eqnarray}
Taking the (discrete) Fourier transform of the linearized hydrodynamic
equations
around HCS, it is easy to check that the transverse velocity perturbation
decouples from the rest and
satisfies the equation
\begin{equation}
\rho_H \frac{\partial \delta{\bf v}_{k_\perp}}{\partial s} = \lambda_k \delta {\bf v}_{k_\perp}
\label{lincrit}
\end{equation}
with
\begin{equation}
\lambda_k \equiv -\frac{4 k^2 \pi^2}{L^2}\eta_0 +
\frac{\rho_H^2 \omega_0}{2} \,\,\,\, ; \,\,\,\, \{ k_x, k_y \} \, = \,0,
\pm 1,\, \pm 2, \, \dots
\end{equation}
For small values of $\omega_0$ (proportional to the dissipation) $\lambda_k
< 0$, so that the
perturbations $\delta {\bf v}_{k_\perp}$ decay exponentially
(the case $k=0$ should not be taken into account since the center of
mass velocity is strictly zero).
But there exists a critical
value of
$\omega_0$ for which
$\lambda_k$  vanishes, thus indicating the stability limit
of the corresponding mode.  The first modes to become unstable
correspond to $|{\bf k}| = 1$
(i.e, $k_x = \pm 1, \, k_y =0$ and $k_x =0, \,k_y = \pm 1 $). The instability
threshold for $\omega_0$ is then given by
\begin{equation}
\widehat{\omega}_0 = \frac{8 \pi^2 \eta_0}{\rho_H^2 L^2}
\label{crpoint}
\end{equation}
The stability for the other modes has been studied previously
\cite{brey96a}. In this last reference, it was shown that for, low
dissipation, the first instability that arises is indeed the transverse
velocity instability.

The origin of the instability can be understood also in terms of the real time
hydrodynamics.  In fact, the relaxation of the transverse hydrodynamic
velocity is basically due to the
viscous diffusion, which depends on the system size, whereas the cooling
process is governed by local
dissipative collisions.   There exists therefore a system length beyond
which the dissipation of thermal
energy is faster that the relaxation of the transverse hydrodynamic
velocity.  The latter will then
increase, when observed on the scale of thermal motion, thus producing the
shearing pattern.


\section{Nonlinear analysis of the instability} \label{nonlinear}

In this section we propose to work out the explicit form of the velocity
field beyond the instability.  The
calculations are tedious and quite lengthy, so that, here, we only describe
explicitely the basic steps. We start by taking the Fourier transform
of the full nonlinear hydrodynamic
equations, obtaining a set of coupled nonlinear equations for the modes
\{ $\delta \rho_k, \delta T_k, \delta {\bf v}_{k_\parallel}, \delta
{\bf v}_{k_\perp}$ \}, where $\delta
{\bf v}_{k_\parallel}$ represents the longitudinal component of the
velocity field. Close to the
instability ($\omega_0
\approx
\widehat{\omega}_0$,
$|{\bf k}| = 1$), the modes
$\delta {\bf v}_{1_\perp}$ exhibit a {\it critical slowing down} since
$\lambda_1 \approx 0$, whereas the other hydrodynamic modes decay
exponentially (cfr. ref.
\cite{brey96a}).  On this slow time scale, i.e.
$s \approx O(\lambda _1^{-1})$, the ``fast'' modes
\{$\delta \rho_k; \delta T_k; \delta {\bf v}_{k_\parallel};\delta
{\bf v}_{k_\perp}, k\neq 1 $\}
can be considered as stationary, their time
dependence arising mainly through
$\delta {\bf v}_{1_\perp}$. Setting the time derivatives of these fast
modes to zero one can
express them in terms of the slow modes $\delta {\bf v}_{1_\perp}$. If now
one  inserts
the so-obtained expressions
into the evolution
equation for the slow modes, one gets a set of closed nonlinear equations
for $\delta {\bf v}_{1_\perp}$ ({\it adiabatic
elimination} \cite{haken,nicolis}), usually referred to as {\it normal
form} or {\it amplitude
equations} [G. Nicolis,  {\it Introduction to Nonlinear Science}, Cambridge
University Press (1995) ].
Note that such a calculation is only possible close to the instability
threshold, where the amplitude of
$\delta {\bf v}_{1_\perp}$ approaches zero as $\omega
\rightarrow \widehat{\omega}_0$.  On the other hand, in this limit the only
Fourier modes that will have
large amplitudes are the transverse velocity modes with wavevector equal to
$\pm \, 2 \pi/L$ (i.e.
$|{\bf k}| = 1$). There are four such modes:
\begin{equation}
\delta {\bf v}_{k_\perp} = \left\{\begin{array}{ll}
A_1             \,\,\,\,          & {\bf k}=(1,0) \\
A_1^*           \,\,\,\,          & {\bf k}=(-1,0) \\
A_2             \,\,\,\,          & {\bf k}=(0,1) \\
A_2^*           \,\,\,\,          & {\bf k}=(0,-1) \\
\end{array}\right.
\end{equation}
\\
After some tedious algebra, one finds
\begin{eqnarray}
\label{eqa1}
\rho_H\frac{dA_1}{ds} & = & \lambda_1 A_1
-4 \pi^2 L^{-2} \eta_0\left( C_1
\left|A_1\right|^2
+ C_2 \left|A_2\right|^2\right) A_1\\
\label{eqa2}
\rho_H\frac{dA_2}{ds} & = & \lambda_1 A_2
-4 \pi^2 L^{-2} \eta_0\left( C_1
\left|A_2\right|^2
+ C_2 \left|A_1\right|^2\right) A_2
\label{EqAmps}
\end{eqnarray}
where
\begin{eqnarray}
C_1 &=& \frac{8(k_0-\mu_0)-3\eta_0}{8(k_0-\mu_0)-2\eta_0}\\
C_2 &=& \frac{2(k_0-\mu_0)+\eta_0}{2(k_0-\mu_0)-\eta_0}
\end{eqnarray}
The amplitude equations (\ref{eqa1}) and (\ref{eqa2}) admit three different
stationary solutions:
\begin{eqnarray}
(a)\hspace{0.3cm}&&A_1=0, \hspace{0.5cm} A_2=0
\label{stat1}\\
(b)\hspace{0.3cm}&&\left| A_1 \right| =\widehat{A} , \hspace{0.5cm} A_2=0
\label{stat2}\\
&& A_1=0 , \hspace{0.5cm} \left| A_2\right|=\widehat{A}
\label{stat3}\\
(c)\hspace{0.3cm}&&\left|A_1\right| = \left|A_2\right|
=\widehat{A} \, \sqrt{\frac{C_1}{C_1 + C_2}}
\label{stat4}
\end{eqnarray}
where we have set
\begin{equation}
\widehat{A}=\frac{L}{2 \pi} \, \sqrt{\frac{\lambda_1}{\eta_0 C_1}}
\end{equation}
Note that the phases of the above stationary solutions are arbitrary
(recall that $A_1$ and $A_{-1}$ are
complex conjugate).

The trivial solution (a) corresponds to a motionless fluid, whereas the
solutions (b) are shearing states with the corresponding fluxes oriented
either in
the $y$ or $x$ direction.  The {\it mixed mode} solution (c) represents a
vortex state with two counter-rotating vortices in the box.

Below the critical point ($\lambda_1 < 0$), the trivial solution (a) is the
only stable one. As we cross
the critical point this solution becomes unstable.  A linear stability
analysis of the Eqs.
\ref{EqAmps} shows that the shearing states are stable provided
\begin{equation}
\frac{C_2}{C_1}>1 \label{coco}
\end{equation}
while the mixed mode solution is unstable.
Satisfying Eq.\ref{coco} depends on the values of the
transport coefficients.  For small dissipativity it is always fulfilled,
provided the number density
remains relatively low. In fact, a mixed mode state has been observed
recently in a highly dense
system
\cite{mcnamara96}.

The occurence of either shearing states depends on the initial
state.  In a
statistical sense, they are equally probable.   For example, in the case
that the system chooses the
solution \{$A_1=\widehat{A}$, $A_2=0$\}, the velocity field reads
\begin{eqnarray}
{\bf v} = 2 \widehat{A} \cos(2 \pi x/L) \, \widehat{\bf y}
\label{vfield}
\end{eqnarray}
where $\widehat{\bf y}$ represents the unit vector in $Y$ direction. The
temperature and density perturbations are
then given by
\begin{eqnarray}
\label{Tfield}
\delta T &=& - (\widehat{A})^2\left[1+(1-C_1) \cos(4 \pi x/L) \right]\\
\delta\rho &=& \rho_H (\widehat{A})^2(1-C_1) \cos(4 \pi x/L)
\label{rhofield}
\end{eqnarray}
The instability of the transverse velocity field thus gives rise to
modifications
of the temperature and density fields. The temperature decreases globally,
since part of the energy is
taken by the convective motion.  Moreover, because of the viscous heating,
the temperature profile exhibits
a spatial modulation, i.e. it is higher where the viscous heating is higher.
The density profile also shows a spatial modulation that keeps the pressure
homogeneous (recall that for a
two dimensional low density  gas,
$\delta p \approx \rho_h \delta T + \delta \rho$, in system units).
This modulation of the density, that was first observed by Goldhirsh
and Zanetti \cite{goldhirsch93} (Fig. 2 of their article) in the study
of the clustering instability, is stable and does not lead to
clustering.

Using the above expressions for the hydrodynamic fields, the energy
density profile reads
\begin{equation}
e  =  \rho_H \left(1+\widehat{A}^2 \cos(4 \pi x/L) \right)
\label{efield}
\end{equation}

Assuming the molecular chaos hypothesis, the mean collision rate,
$\overline{\nu}$, and  dissipation rate,
$\overline{\omega}$, can be computed as well:
\begin{eqnarray}
\overline{\nu} &=& \frac{2 \sqrt{\pi}}{V \rho_H} \int \rho^2 \sqrt{T}
\, dV \,\, =
\,\, \nu_0\rho_H
\left(1-\frac{\widehat{A}^2}{2} \right)
\label{nuteor}\\
\overline{\omega} &=& \frac{1}{V}\int \omega \, dV  \,\, = \,\,  \omega_0
\rho_H^2 \left(1-\frac{3
\widehat{A}^2}{2}
\right)
\label{omegateor}
\end{eqnarray}
These relations show that the global decrease of the
temperature leads to corresponding decreases of the
collision frequency and dissipation rate.

It is important to note that the origin of the nonlinear coupling of slow
modes lies in the viscous
heating term and the state-dependence of the transport coefficients, and
not in the usual convective
derivatives ($({\bf v} \cdot \nabla) {\bf v}$). In fact, as we have shown
above, the
shearing state produces a variation in the density and temperature fields
that modify locally the value of
the transport coefficients. This effect, which is negligible in classical
fluids, can become very important
in granular fluids, mainly because of the the lack of scale separation of
the kinetic and hydrodynamic
regimes. Contrary to normal fluids, here, the convective energy is
comparable to the thermal
energy. In fact, as will be shown in the MD simulations, in a well
developed shearing state up to half
of the total kinetic energy corresponds to the convective motion.  The
microscopic source of this phenomenon
lies in the fact that only the relative energy is dissipated in binary
collisions, i.e. the center of mass
energy is conserved. In other words, the thermal energy is dissipated but
the convective one is conserved.
This asymmetry in the energy dissipation mechanism is at the very origin of
the shearing
instability.


\section{Molecular dynamics simulations} \label{mdsim}

For the molecular dynamic simulations we have considered a system made
of $N\!=\!10000$
hard disks with a global number density $\rho_H\!=\!0.005$.  Inelastic
collision rules are adopted,
with a dissipativity varying from
$q\!=\!0.0$ to $q\!=\!0.12$. The boundary conditions are periodic in all
directions. A spatially homogeneous initial condition is adopted, with
velocities sorted from an
equilibrium (zero mean velocity) Maxwellian distribution.  We note that the
density is low enough so
that the system remains within the low-density regime.

The simulations have been performed in the rescaled time. Computationally,
this is achieved by doing a normal IHS simulation (event driven molecular
dynamics \cite{MRC}), but at each collision the value of the kinetic energy
is updated according to the energy dissipated. The instantaneous value of
$\gamma$ is computed from the kinetic energy, allowing to evaluate all the
rescaled quantities. Note that the $s$-time can be integrated in the
simulation because $\gamma$ is a piecewise constant function, thus allowing
to make periodic measurements in the system.   Finally, to avoid roundoff
errors, a real velocity rescaling is performed whenever the kinetic energy
decreases by a given amount  (typically $10^{-7}$ of the initial value).

\begin{figure}[htb]
\epsfig{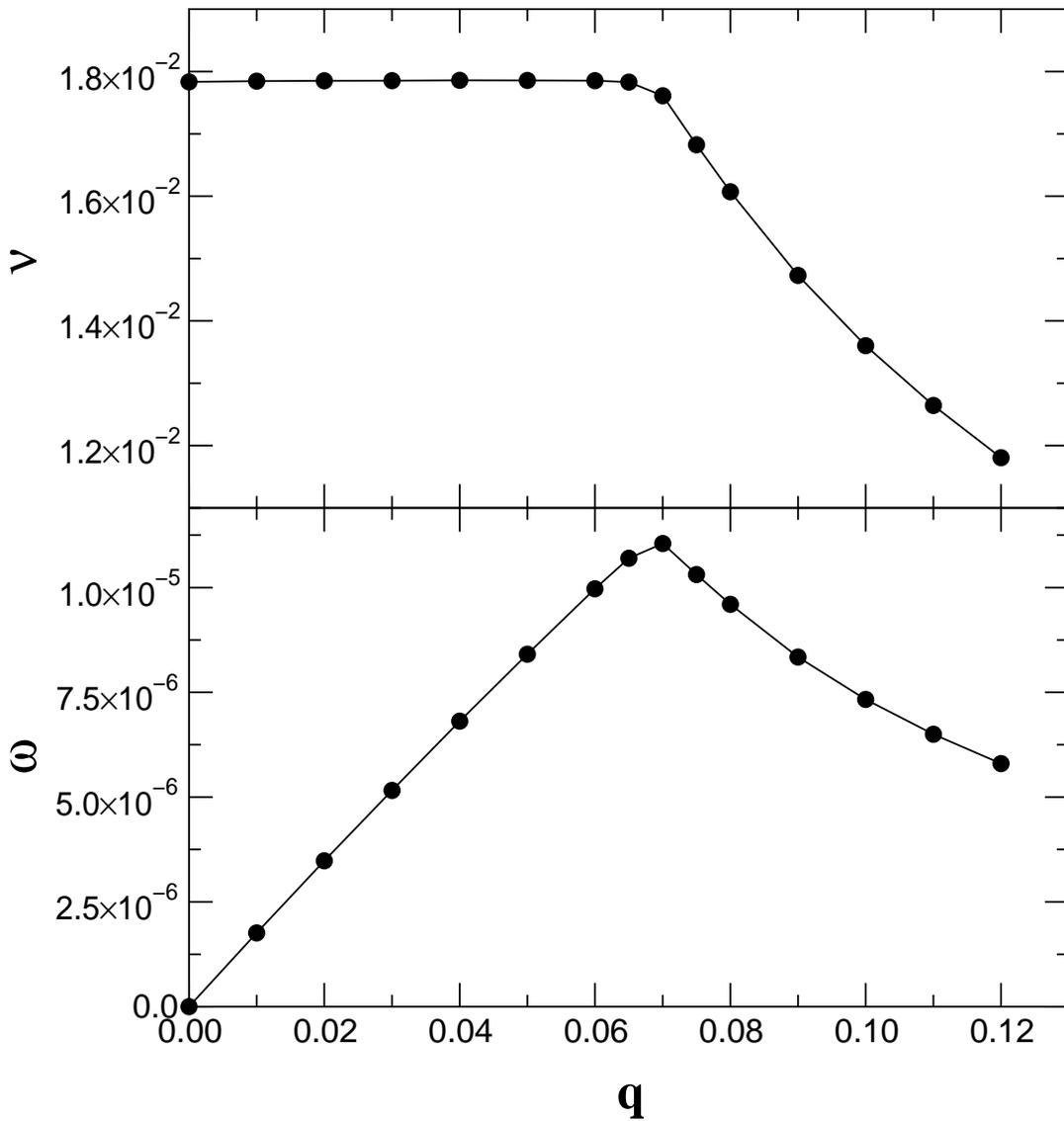}
\caption{Mean collision frequency $\overline{\nu}$ and dissipation rate
$\overline{\omega}$ as a function of the
dissipativity $q$ obtained in molecular dynamics simulations with $N=10000$
particles and density $n=0.005$. The lines are drawn for easier
reading of the graphs.}
\label{nusim}
\end{figure}

In each simulation, the collision frequency and temperature dissipation
rate are computed with respect to the $s$-time, after the system has
reached a stationary regime. In
Fig. \ref{nusim}, the collision frequency and the temperature dissipation
rate are presented as a
function of the dissipation coefficient $q$.  As expected, the shearing
instability
is associated with an abrupt decrease of these functions.
It must be noted that the decreasing of the collision frequency is more
that $30\%$, which corresponds to a global decrease of the temperature of more
than $50\%$. This means that when the shearing is fully developed, about
half
the total kinetic energy is taken by the macroscopic motion. This phenomenon is
typical of granular media and has no counterpart in classical fluids.

To measure the critical point, $q_0$, we fit the collision frequency 
according to the following piecewise function\begin{equation}
\overline{\nu} = \left\{ \begin{array}{ll}
a_0 & q\leq q_0\\
a_0+a_1(q-q_0) + a_2(q-q_0)^2 & q>q_0
\end{array}\right.
\end{equation}
obtaining
\begin{eqnarray}
q_0 &=& 0.0686 \\
a_0 &=& 0.0178\\
a_1 &=&-0.167\\
a_2 &=& 1.03
\end{eqnarray}

Similarly, the dissipation rate is fitted according to
\begin{equation}
\overline{\omega} = \left\{ \begin{array}{ll}
b_0 q & q\leq q_0\\
b_0 q +b_1(q-q_0) q+ b_2(q-q_0)^2 q & q>q_0
\end{array}\right.
\end{equation}

Using $q_0=0.0686$, one finds
\begin{eqnarray}
b_0 &=& 0.000166 \\
b_1 &=&-0.00466\\
b_2 &=& 0.0546
\end{eqnarray}

To compare these results with our theoretical predictions we need the
explicit form of the transport coefficients, up to critical
dissipativity.  Unfortunately, there are no known expressions for them in
the case the $2d$ IHS model in the low-density regime. However, as the
critical dissipativity is small we can use the the quasielastic approximation
for the transport coefficients (that is, taking the first non-trivial order
in $q$)
\cite{sengers66,jenkins85c}
\begin{eqnarray}
\omega_0 &=& 4 \sqrt{\pi} q\\
\eta_0   &=& \frac{1}{2\sqrt{\pi}} \\
k_0      &=& 2/\sqrt{\pi} \\
\mu_0    &=& 0
\end{eqnarray}

The critical dissipativity and the amplitude of the shearing state are
given by
\begin{eqnarray}
\widehat{q} &=& \frac{\pi}{\rho_H^2 L^2}
\label{qteor}\\
\widehat{A} &=& \rho_H L \sqrt{\frac{30}{29 \pi} \delta q}\\
      &=& \sqrt{\frac{30}{29} \frac{\delta q }{\widehat{q}}}
\label{Ateor}
\end{eqnarray}

For the presented simulation, the predicted critical dissipativity is
\begin{equation}
\widehat{q}= 0.0628
\end{equation}
which shows a discrepancy of $8\%$ with the observed value. This difference
is consistent with the adopted approximations.

The predicted values for $a_0$, $a_1$, $b_0$, $b_1$ (cfr. Eqs.
\ref{nuteor}, \ref{omegateor}, \ref{qteor}, and \ref{Ateor}) are

\begin{eqnarray}
a_0 &=&0.0177\\
a_1 &=&-0.146\\
b_0 &=&0.000177\\
b_1 &=&-0.00438
\end{eqnarray}
which are also consistent with the adopted approximations.

Since the system is periodic, the developed convective pattern can diffuse
in the direction perpendicular to the flow (the phases of the complex
amplitudes $A_i$ are arbitrary due to Galilean invariance).  As a
result, the average hydrodynamic fields remain vanishingly small, mainly
because of ``destructive'' interference. To overcome this difficulty we
have performed another series of simulations, keeping periodic boundary
conditions in the vertical direction, while introducing a pair of
stress-free and perfectly insulating parallel walls in the horizontal
direction (in a collision with a wall the tangential velocity is conserved
whereas the normal one is inverted).  As a consequence, the total vertical
momentum is conserved, which will be simply set to zero initially.

The nonlinear analysis for this case is similar to the periodic one, except
that, here, the direction of the flow pattern remains always parrallel to the
walls.  Furthermore, the
unstable wavevector is now $k=\pi/L$,  because of to the fixed boundary
conditions. As a result all the
previous predictions remain valid, except that everywhere $L$ must be
replaced by $2 L$.

We have used the very same number of particles and
density for this series of simulations, but, of course, the different boundary
conditions produce a new critical dissipativity. Performing the same
analysis as before, the
measured critical point and fit parameters turn out to be
\begin{eqnarray}
q_0 &=& 0.0163\\
a_0 &=& 0.0178\\
a_1 &=&-0.562\\
a_2 &=& 8.93\\
b_0 &=& 0.000175\\
b_1 &=&-0.0179\\
b_2 &=& 1.06
\end{eqnarray}

while the predicted ones are
\begin{eqnarray}
\widehat{q} &=& 0.0157\\
a_0  &=&0.0177\\
a_1  &=&-0.583\\
b_0 &=&0.000177\\
b_1  &=&-0.0175
\end{eqnarray}

\begin{figure}[htb]
\epsfig{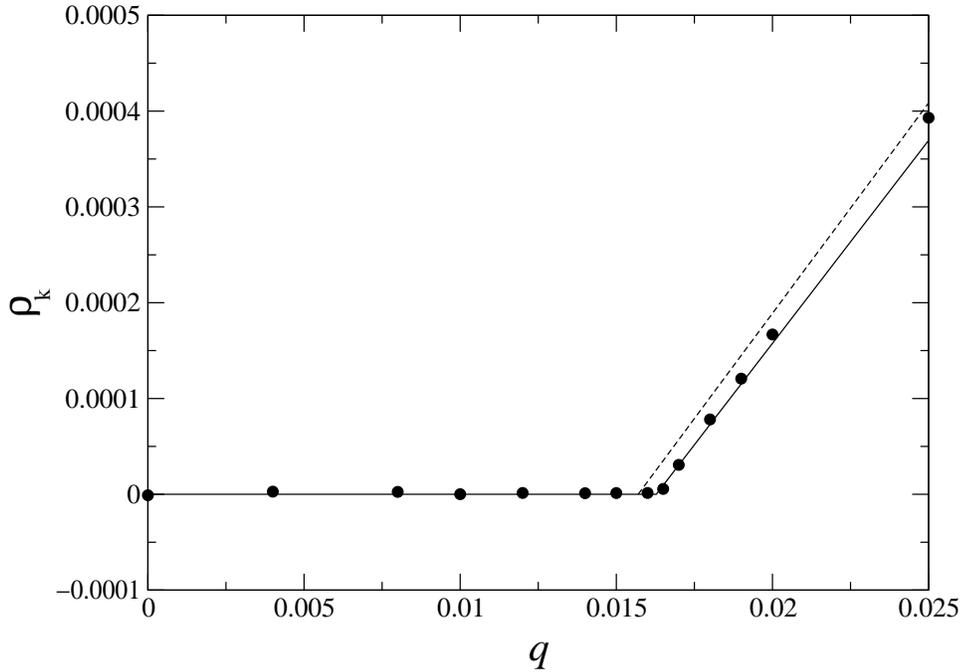}
\caption{Amplitude of the $k_x=2 \pi/L$ cosine component of the density field
as a function of the dissipativity $q$.
The dots are the results of molecular dynamics simulations, the dashed line
is the full nonlinear theory prediction, and the solid line is the nonlinear
theory prediction where the critical point is taken from the fit. 
The simulations were done in a channel with
$N=10000$ particles and density $n=0.005$.}
\label{camposampn}
\end{figure}

\begin{figure}[htb]
\epsfig{file=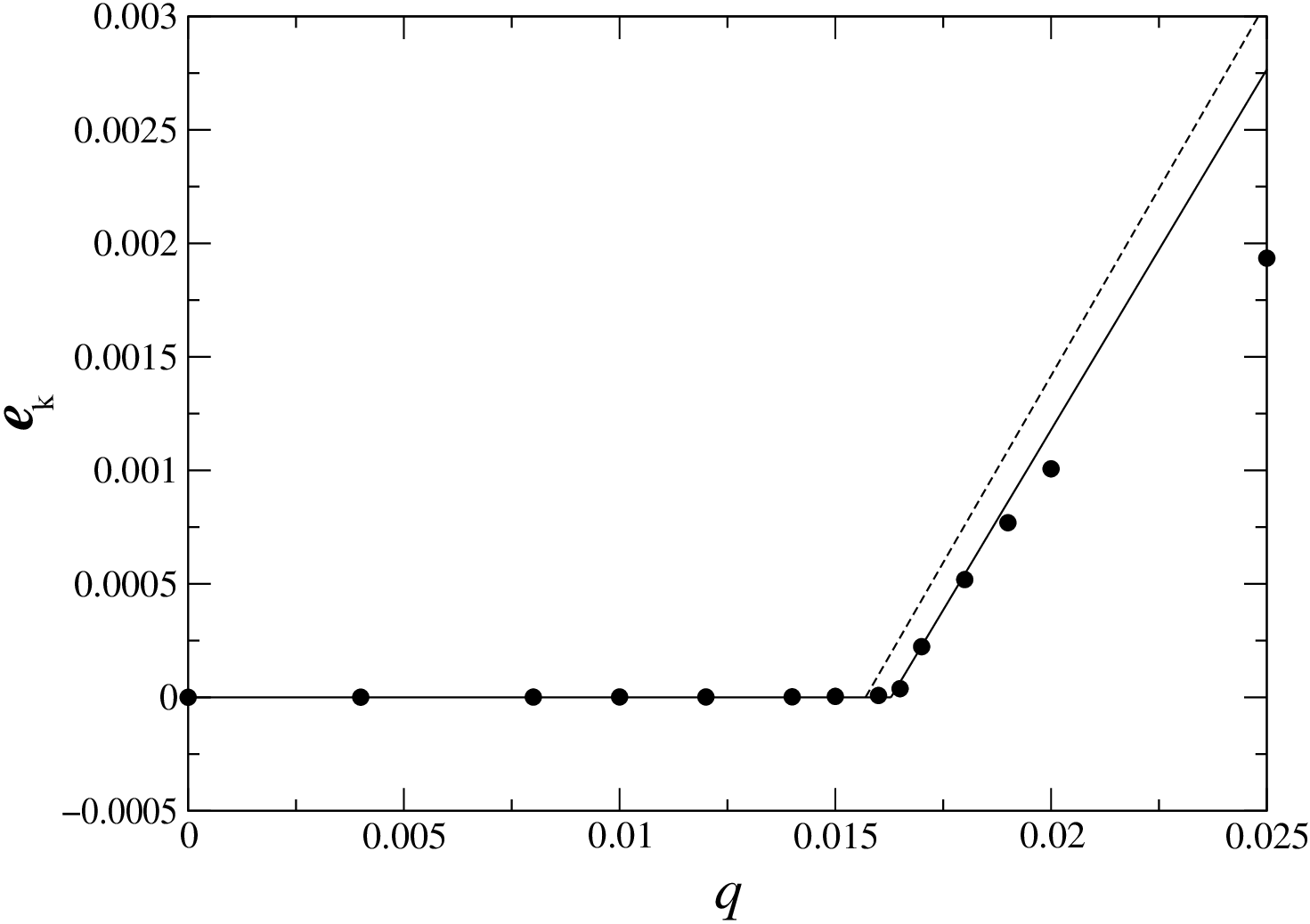,width=3.5in,angle=270}
\caption{Amplitude of the $k_x=2 \pi/L$ cosine component of the energy density
field as a function of the dissipativity $q$. Graph symbols and simulation
parameters are explained in Fig. \protect{\ref{camposampn}}.}
\label{camposampe}
\end{figure}

\begin{figure}[htb]
\epsfig{file=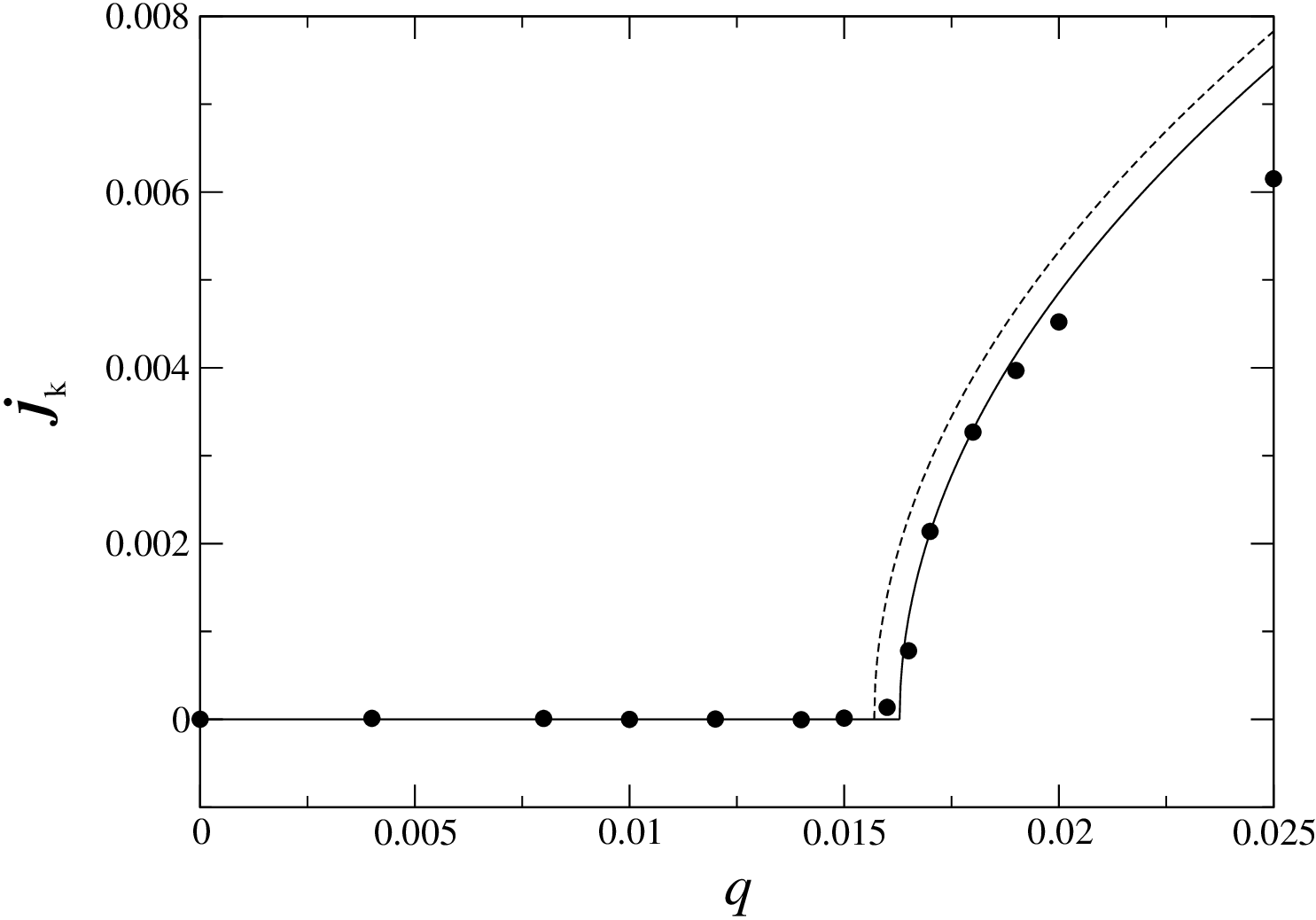,width=3.5in,angle=270}
\caption{Amplitude of the $k_x=\pi/L$ cosine component of the momentum density
field as a function of the dissipativity $q$. Graph symbols and simulation
parameters are explained in Fig. \protect{\ref{camposampn}}.}
\label{camposampj}
\end{figure}

Eqs. \ref{vfield}, \ref{rhofield}, and \ref{efield}, once $L$ is
replaced by 
$2L$, indicate that the perturbation of the transverse momentum density
(${\bf j}=\rho{\bf v}$) has a wavevector
equal to $k_x=\pi /L$,
while the density and
energy density have wavevectors
$k_x=2\pi/L$.
In the simulations we computed the amplitudes of these Fourier
modes using the microscopic definitions for the particle, momentum, and
energy densities.
Figs. \ref{camposampn}, \ref{camposampe}, and \ref{camposampj} show
the predicted and computed Fourier modes amplitudes.
The predictions are in good agreement with the simulations in the
neighborhood of the critical point, showing that not only the average
quantities like the collision frequency are well predicted, but also the whole
hydrodynamic picture is correct.

\section{Conclusions} \label{conclusion}

Taking advantage of the lack of energy scale in the IHS model, a
rescaled time formalism was introduced that allows us to study the
homogeneous cooling state as a nonequilibrium steady state.
Using a hydrodynamic description for granular media written with a
rescaled time variable, the
shearing instability has been studied in the nonlinear regime. It has been
shown that the shearing state is the stable solution and its amplitude
has been computed. The appearance of the velocity field produces
that part of the kinetic energy goes from the kinetic to the hydrodynamic
scale. In usual fluids this redistribution of the energy is
negligible, but in granular fluids it can represent an important fraction
of the total energy.  This phenomenon is a manifestation of a global
property of granular fluids: there is not in general a clear separation
between the kinetic and the hydrodynamic regimes. This could lead to
put into question the validity of a hydrodynamic description.
Nevertheless, at small values of the dissipativity coefficient,
predictions based on the nonlinear hydrodynamic equations are in
excellent agreement with molecular dynamics simulations. Both the
value of the critical dissipativity and the behavior after the
instability has developed are well predicted. This is a remarkable result
which shows again how robust the hydrodynamic fluid equations are when
they are tested at time and length scales where their validity could be
questioned.

\acknowledgments{This work is supported by a European Commission DG 12
Grant PSS*1045 and by a grant from FNRS Belgium. One of us (R.S.)
acknowledges the grant from {\em MIDEPLAN}.}



\clearpage

\clearpage

\end{document}